\documentclass[conference]{IEEEtran}
\IEEEoverridecommandlockouts
\usepackage{cite}
\usepackage{amsmath,amssymb,amsfonts}
\usepackage{algorithmic}
\usepackage{graphicx}
\usepackage{textcomp}
\usepackage{xcolor}
\usepackage{xurl}
\usepackage{float}
\usepackage{optidef}
\def\BibTeX{\rm B\kern-.05em{\sc i\kern-.025em b}\kern-.08em
    T\kern-.1667em\lower.7ex\hbox{E}\kern-.125emX}
\begin{document}

\title{Electric Vehicle Aggregation Review: Benefits and Vulnerabilities of Managing a Growing EV Fleet\\
\thanks{A part of this research was supported by the Visiting Faculty Research
Program of the Air Force Labs, Rome, New York. Cleared for publication
Case Number:AFRL-2023-5334}}

\author{\IEEEauthorblockN{Kelsey Nelson${^a}$, Javad Mohammadi${^a}$, Yu Chen${^b}$, Erik Blasch${^c}$, Alex Aved${^c}$, David Ferris${^c}$, \\Erika Ardiles Cruz${^c}$, Philip Morrone${^c}$}
\IEEEauthorblockA{${^a}$Civil, Architectural and Env. Eng., The University of Texas at Austin, Austin, TX 78705, USA \\
$^{b}$Dept. of Electrical \& Computer Engineering, Binghamton University, Binghamton, NY 13902, USA\\
$^{c}$The U.S. Air Force Research Laboratory, Rome, NY 13441, USA\\
\{kelseynelson, javadm\}@utexas.edu, ychen@binghamton.edu, \\ \{erik.blasch.1,alexander.aved, david.ferris.3, erika.ardiles-cruz, philip.morrone.6\}@us.af.mil}
}







\maketitle

\begin{abstract} 
Electric vehicles (EVs) are becoming more popular within the United States, making up an increasingly large portion of the US's electricity consumption. Hence, there is much attention has been directed on how to manage EVs within the power sector. A well-investigated strategy for managing the increase in electricity demand from EV charging is aggregation, which allows for an intermediary to manage electricity flow between EV owners and their utilities. When implemented effectively, EV aggregation provides key benefits to power grids by relieving electrical loads.. These benefits are aggregation's ability to shift EV loads to peak shave, which often leads to lower emissions, electricity generation prices, and consumer costs depending on the penetration levels of non-dispatchable electricity sources. This review seeks to appropriately highlight the broad vulnerabilities of EV aggregation alongside its benefits, namely those regarding battery degradation, rebound peaks, and cybersecurity. The holistic overview of EV aggregation provides comparisons that balance expectations with realistic performance.

\end{abstract}

\section{Background and Motivation}

Electric vehicle (EV) aggregation is the management of an EV fleet by an intermediary that coordinates services between EV drivers and their electricity providers \cite{nelson2023evs}. The intermediary’s role can take different forms and often relies on the use of smart electric vehicle supply equipment (EVSE), which for this review will refer to EV charging infrastructure that is able to send electricity to the grid via bidirectional charging and/or receives direct communication between aggregators and EV drivers. Such communication allows for aggregators to shift loads \cite{DOE2016, mohammadi2023towards, MohammadBOOKCHAPTER} and in some cases coordinate vehicle-to-grid (V2G) services \cite{Scorecard} in order to benefit the grid. Because the EV load is so flexible, aggregation can allow the fleet to charge at times that are conducive to the goals being prioritized, typically (1) to provide peak shaving, (2) to lower a grid's overall emissions and (3) to lower the price of electricity generation. Despite these advantages, EV aggregation has notable drawbacks that must be considered. 

\begin{figure}[htbp]
\centering
\includegraphics[scale=.5]{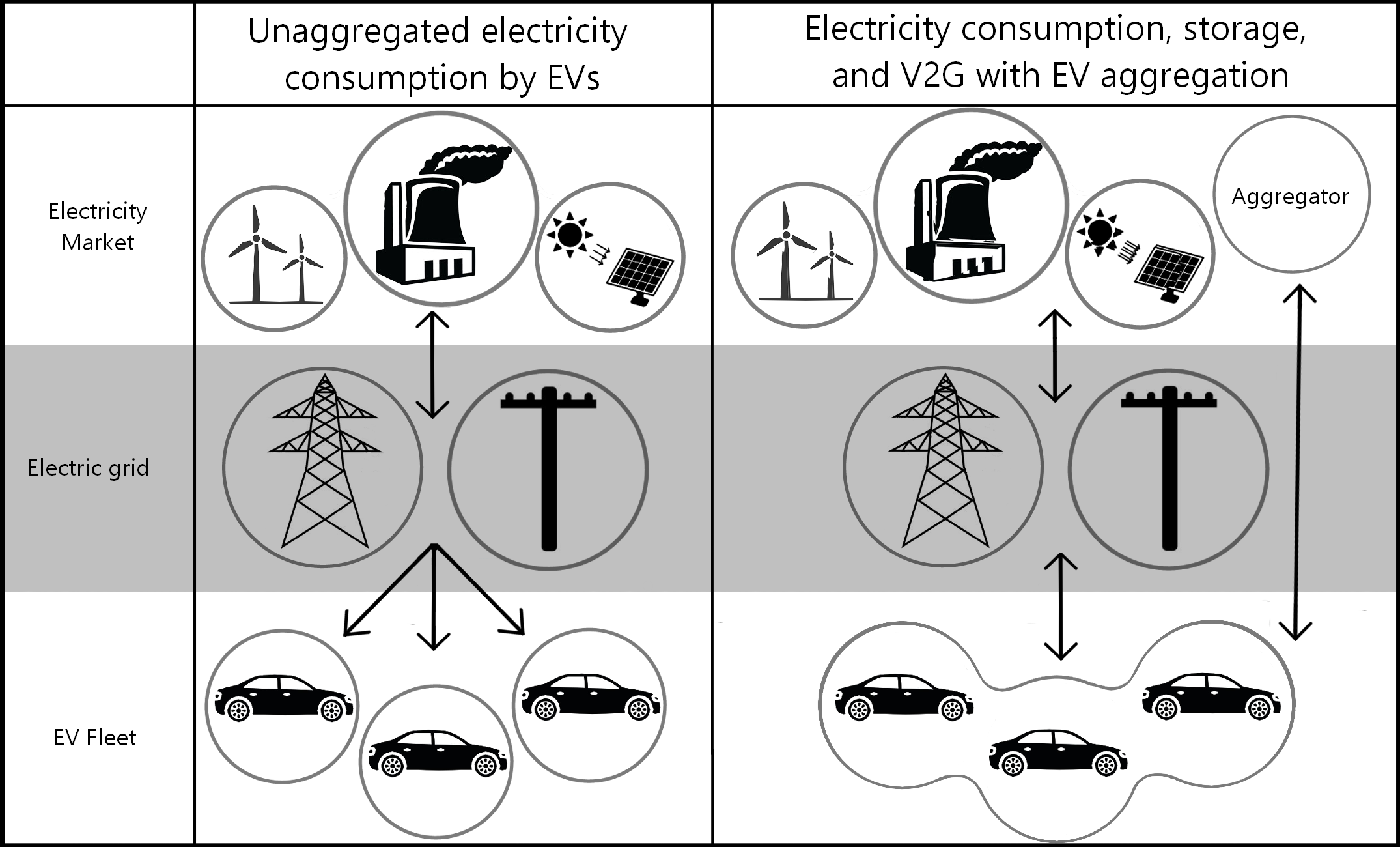}
\caption{Aggregators facilitate bi-directional communication and electricity flow between system components}
\label{DescriptiveFigure}
\vspace{-.2cm}
\end{figure}


Most reviews on the anticipated interactions between EVs and electric grids focus solely on the benefits of EV aggregation. While there are some reviews that do take into account potential pitfalls surrounding EVs at scale, to our knowledge these reviews could be improved through expanded considerations.. Authors in \cite{CostReview} investigate important costs of aggregation, but its scope deals almost strictly with quantifiable financial costs, such as the costs to implement technology for physical sensing, communication avenues, enablement costs, and how these costs are passed onto ratepayers. Reference \cite{PositiveAndNegative} discusses both the positive and negative aspects of EVs on charging systems, but not specifically the implementation of aggregation. Because \cite{PositiveAndNegative} covers solely EV presence, it does not take into account issues unique to EV aggregation, such as price signaling's ability to cause rebound peak, the cybersecurity issues that are inherent to the need for communication via EVSE, and vehicle-to-grid (V2G)'s impact on EV battery degradation, though it lists some of these factors as notable for future research. Lastly, \cite{AggModeling} is a review specifically regarding EV aggregation, however, it is primarily meant to inform aggregators and participating utilities of options for modeling methods for optimal bidding strategy and effective control methods for power systems planning. The authors of this study similarly note that future works should consider trade-offs between different objectives such as energy prices and battery degradation. 

The review in this paper fills a research gap that deals with the multidisciplinary aspects of EV aggregation in order to allow for the informed weighing of both positive and negative externalities of different aggregation strategies. The evaluation considerations emerge from considering a variety of literature spanning different perspectives, such as 
economics, power systems planning, and policy framework. Fig.~\ref{DescriptiveFigure} provides a visualization of how these perspectives come together to form the scope of this review.

\begin{figure}[htbp]
\centering
\includegraphics[scale=.9]{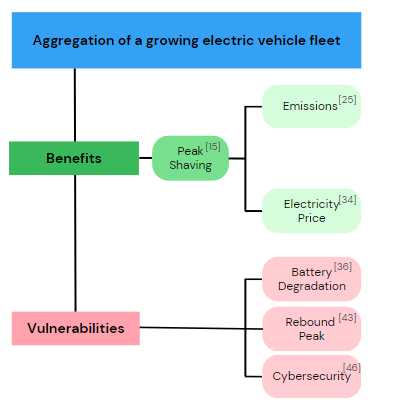}
\caption{EV Aggregation benefits and vulnerabilities covered in this review}
\label{DescriptiveFigure}
\end{figure}

The rest of the paper begins with the known benefits of EV aggregation are summarized, most notably its potential for peak shaving, lowering grid emissions, and lowering electricity prices. Next, it examines the downsides of EV aggregation, which are (1) the increase in battery degradation from the use of bidirectional charging (2) a rebound peak effect, where once charging is no longer discouraged during the grid's peak demand, an unwanted secondary peak occurs afterward as several EV drivers plug their vehicles in, and (3) the cybersecurity vulnerabilities that the use of smart EVSE opens users and grids up to. Section \ref{discussion} discusses the extent to which these externalities would impact the different stakeholders within the EV space; EV owners, utilities, aggregators, and policy makers.

\section{Benefits}

This section discusses the proven benefits of EV aggregation, providing evidence from real-world implementation as well as citing studies that test different strategies for maximizing aggregation benefits depending on the goal(s) being prioritized. Key to the discussion is peak shaving, which often inherently leads to lower electricity generation and charging prices, and lower grid emissions\cite{PeakShavingBenefits}. 

\subsection{Peak Shaving}

One of the most commonly cited benefits of EV aggregation is its ability to shift EV loads to shave down what the grid's peak power would have been in the absence of charging management \cite{PeakShavingReview}. 

There are several peak shaving strategies that have been investigated and found to be effective. Different control algorithms for both unidirectional charging and bi-directional charging have been thoroughly investigated, such as those that employ model predictive control \cite{MPC}, optimal control \cite{OptimalControl}, and Kalman filtering \cite{KalmanFiltering}. More recently, neural networks and predictive models have been trained and tested using previous user data. \cite{ANN} uses artificial neural networks to provide state of charge (SOC) estimation for aggregators, and \cite{ParkingGarageCase} uses a neural network trained on 2021 data and tested with 2022 data in order to use day-ahead forecasting to implement rule-based peak shaving control.

\subsubsection{Emissions}

Presently, peak shaving often translates directly to a reduction in the carbon intensity of the grid. To serve unusually high levels of demand, peaking power plants must come online, which have exceptionally high carbon intensity \cite{Peakers}. However, many regions of the United States are rapidly shifting their energy mixes by expanding renewable energy source (RES) capacity in order to reach its decarbonization goals \cite{CarbonTarget}. RES balancing means that there will soon be times when RES output may exceed peak demand, making it potentially advantageous in the future to shift the flexible EV load to these times of high demand if the renewable output is also expected to be high \cite{NRELCurtailment}. The electrical-temporal shift would prevent the curtailment of RES output, ensuring that as much electricity consumption is from emission-free sources as possible. 

In \cite{SolarCurtailment}, the authors investigate strategies to best avoid the curtailment of solar output under a high EV adoption and solar penetration scenario. They find that workplace charging along with accelerated EV adoption will be critical to minimizing solar curtailment, with workplace charging having the ability to reduce peak demand while also cutting RES curtailment by as much as 50\%. Similarly, in \cite{WindCurtailment} the authors propose a two-stage peak-shaving strategy using battery energy storage systems (BESSs) in order to create an optimization model with the minimization of wind curtailment as part of its objective function. The problem is solved by a neural network algorithm using data from a real-world case study and is able to reduce wind curtailment by over 40\%.

\subsubsection{Electricity Price}

Limiting the curtailment of emissions-free, non-dispatchable electricity sources (most commonly wind and solar) often inherently lowers the levelized cost of energy (LCOE) for energy providers, due to the fact that these sources are cheaper on average than their fossil fuel counterparts \cite{LCOE}. 

Though there is a strong correlation between RES supply and generation price, electricity price is still influenced by other factors, such as the need to quickly ramp up and down generation to meet highly variable demand. Therefore, it is still important to consider both factors independently in EV charging models which seek to achieve these goals. For example, previously discussed papers, \cite{SolarCurtailment} and \cite{WindCurtailment}, also include the cost within their optimization problems' objective function and find that electricity cost can easily be accounted for while also successfully limiting RES curtailment.  


There is also extensive literature that primarily focuses on the ability of EV aggregation to lower electricity prices without taking emissions into account. For example, \cite{URREHMAN2022108090} uses an optimal hierarchical aggregation algorithm for using V2G for day-ahead charging cost minimization, \cite{WU2021106808} employs a probability prediction model for estimating driver behavior for a day-ahead charging model, and \cite{LoadLevelingAndCost} uses a price-based demand response model in three separate aggregation case studies in order to stabilize the grid via load leveling while minimizing electricity generation and EV charging costs.

\section{Vulnerabilities}
\label{sec:Vulnerabilities}

Though the benefits from EV aggregation are numerous, they are not without negative consequences. As EV adoption rates continue to grow and light duty vehicles (LDVs) become increasingly electrified, awareness of the potential drawbacks and externalities of EV aggregation will be key to making informed decisions in order to maximize its beneficial effects. It's also key to understanding both the public's and utility's willingness to participate, as they can be directly affected by key vulnerabilities. The vulnerabilities that will be discussed in this section are battery degradation, a rebound peak effect, and cybersecurity issues. 

\subsection{Battery Degradation} 

Currently, commercial EVs are manufactured and sold with Lithium-ion battery (LIB) packs, which degrade naturally over their lifetime. However, there is evidence that suggests that aggregation methods can accelerate this process. This directly impacting EV owners by decreasing their battery's range via capacity fade, and in some cases necessitating a costly battery replacement.

Aggregation often leads to increased battery degradation because charging patterns affect how a battery degrades. Charging patterns which lead to lower electricity generation prices are often not the same charging patterns that preserve the battery's longevity \cite{BatteryDegConditions}. Furthermore, this degradation is compounded when directional charging is implemented \cite{BatteryDegV2G}.

The findings from \cite{BatteryDegV2G2} state that degradation from V2G is negligible if V2G is used fewer than 20 times per year. However, states with extreme weather (e.g. extremely high or low temperatures which increase demand for electricity use), such as Arizona and Texas have far more than 20 days per year of peaking events \cite{DallasWeather}, \cite{PheonixWeather}, meaning that it may not be reasonable to assume that V2G would be used within \cite{BatteryDegV2G2}'s threshold for minimal degradation. In order to quantify this risk in scalable monetary terms, they assign an approximate battery degradation cost to the user of \$.38 per 2-hour event using an at-home level 1 charger and \$.82 per 2-hour event using a level 2 charger.

\subsection{Rebound Peak}

Paradoxically, though one of the main benefits of EV aggregation is its potential to limit peak demand, attempts to suppress EV charging during times of high demand can indirectly lead to the creation of a secondary peak later in the day, which is often referred to as a ``rebound peak". 

In \cite{IETRebound}, the likelihood of oscillatory behavior from charging with both price-based and load-based energy management systems (EMS) is showcased. The authors use real Australian household energy consumption data and weather patterns in a hypothetical microgrid with high PV penetration to obtain their results, making the study especially timely as many countries, including the United States, expand both residential and utility-scale solar \cite{SolarProjections}. Authors in\cite{OvernightManagement} provide a case study for conventional overnight charging management strategies during the winter season in Quebec. It confirms the anticipated rebound effect in a local EV fleet and presents a forecasting-based framework with historical charging profiles in order to mitigate this effect. The framework was successful in lowering peak-to-off-peak ratios from a baseline case by 21\% with a noted “very low” cost of implementation. Similarly, \cite{SummerRebound} finds that discouraging charging during their study's system peak hours, which occurred in early summer evenings, caused a rebound peak once charging was no longer discouraged.

\subsection{Cybersecurity}

Because smart EVSE must be interconnected as an Internet of Things (IoT) device in order to allow for effective communication between aggregators, utilities, and other participating agents; the connectivity would increase exploitable vulnerabilities, increasing the risk of cyberattacks, such as those involving grid disruption, impersonation, and data breaches \cite{CybersecurityReview}. 


There are many different avenues that attackers may take in order to disrupt grid operations. In \cite{osti}, several different scenarios for distributed denial-of-service (DDoS) attacks are modeled at an aggregated level in order to see the end result that coordinated malicious attacks could have on the grid. The authors' simulated results found that using DDoS for a sudden load drop yielded minimal change to the grid's frequency. Additionally, load modulation to target the grid's resonant frequencies was found to have negligible impacts on operations. Furthermore, in simulating oscillations on the distribution system, they found that there is ``virtually no risk" of distributed energy resource equipment tripping from such a load modulation event. The authors also note that they simulated an ``extreme case" where all smart EVSE in the region are in use simultaneously during a peaking event, and all stations are at the end of the feeder. In \cite{V2GInjection}, load manipulation attacks are simulated but with the introduction of power injections via V2G. Its findings state that even at low EV penetration levels, the attacker could disrupt grid operations even if outages do not occur. Additionally, the authors find that it would be plausible to use EV attacks in order to hide transmission outages by falsifying data in order to manipulate the reported load at both ends of a disconnected line to grid operators. Lastly, \cite{INL} finds that at the device level, load modulation is able to produce harmonic distortion of over 20\% and reduce the EVSE's power factor to below .8. However, at the grid level, it found that current penetration levels of EVSE are insufficient for attackers to be able to trigger outages.

Outside of direct grid and device disruption, \cite{Impersonation1} models impersonation attacks on advanced, hypothetical GPS-based wireless EV charging networks that use mobile energy disseminators (MEDs). It is shown that an attacker could effectively use GPS spoofing to cut in line in a charging queue, disturbing the optimization of MED routing within an EV network and increasing average travel time \cite{Impersonation1}. 

Disclosure attacks, which seek to surpass authorization to obtain data, are difficult to directly model. Important software and tools used in industry to identify weak points for disclosure attacks, however, have been discussed in published literature. For example, \cite{osti} discusses STRIDE (Spoofing, Tampering, Repudiation, Information disclosure, Denial of service, and Elevation of privilege). Similarly, the Unified Modeling Language for Secure Systems Development was used in \cite{INL}. There have also been well-documented examples of successfully executed disclosure attacks within broad EV systems, such as Tesla's 2023 data breach \cite{TESLA}.

Given current penetration levels within the United States for both EVs and smart EVSE, it is highly unlikely that coordinated cyberattacks involving frequency or load manipulation would have substantial, system-wide impacts on the grids that they target. Additionally, many power grid operators still prefer robustness over autonomous decision-making capability. However, disclosure attacks have already been carried out both within the industry and in controlled simulations via eavesdropping \cite{Eavesdropper}. Additionally, with EV adoption rates on the rise, large-scale plans to expand the US public charging network already in place \cite{TXDOT}, and utility incentives for the installation of EV chargers \cite{UtilityIncentives} grid impacts will move more to the forefront of cybersecurity concerns within the EV space, meaning that aggregators will need to be careful that their serviced networks are following cybersecurity compliance guidelines using robust hardware and defense strategies \cite{ParticleGuidelines}.

\section{Discussion}
\label{discussion}
\subsection{Stakeholder Trade-offs}

\begin{figure}[htbp]
\centering
\includegraphics[scale = .6]{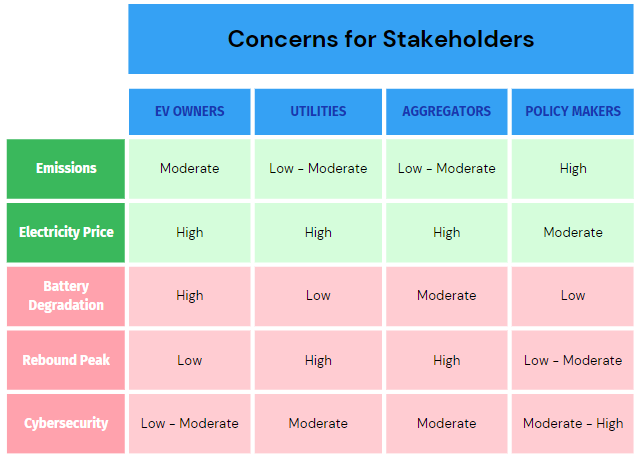}
\caption{Stakeholders within the EV aggregation space and their anticipated concern levels with the benefits and vulnerabilities of EV aggregation}
\label{Stakeholders}
\vspace{-.4cm}
\end{figure}

Each of the discussed benefits and vulnerabilities will be differently prioritized by stakeholders within the EV aggregation space. The key identified stakeholders that have been mentioned so far within this paper are EV owners, utilities, aggregators or fleet managers, and policymakers. This section will discuss their likely concern levels with the benefits and vulnerabilities of aggregation that are covered in this paper, as summarized by Fig.~\ref{Stakeholders}. The concern levels are determined by considering the likelihood of the stakeholder experiencing the consequences of a given externality, the severity of such consequences, their awareness of the externality, and the availability of preventative measures.

\subsubsection{EV Owners}

EV owners also serve the role of electricity consumers within power systems, who are known to be largely motivated by price \cite{PriceResponse}. Though the US consumer base reports high levels of concern regarding sustainability when it comes to purchasing decisions \cite{SustainabilityPrioritization}, many find the price to be an inhibiting factor when it comes to prioritizing sustainability \cite{SustainabilityCost}. Given the prevailing preference for low prices, electricity price is categorized as a high concern with emissions only being categorized as moderate.

Battery degradation is a direct and proven cost incurred by EV owners who participate in aggregation, so in a similar manner to electricity price, it is also expected to be of high concern. Regarding rebound peak, it is assumed that the average EV owner is not aware of how price signaling can cause oscillatory charging behavior that can lead to an unintentional rebound peak, making it a low concern for these stakeholders. Lastly, while the types of cyberattacks that would impact EV owners (disclosure, DDoS) are currently isolated to very rare occurrences, the issue has been gaining more attention lately in mainstream media \cite{CyberSecInMedia1}, \cite{TESLA}, \cite{CyberSecInMedia2}, so it was placed in a category of low to moderate concern.

\subsubsection{Utilities}
\label{subsubsec:Utilities}

In the US, the largest portion of the average utility's customer base is residential/individual consumers, closely tying their concerns as stakeholders with those of the EV owners they service  \cite{UtilityBreakdown}. As most of the utility market is served by for-profit companies \cite{UtilityMarket}, however, they will largely inherently prioritize maintaining competitive electricity rates, keeping electricity cost as a high concern \cite{EnvironmentalEconomics}. While utility concern for emissions has historically been low, differences in regional customer base ideology and state incentives may cause some utilities to heighten their concern with emissions, so it was categorized as being of low to moderate level. 

Battery degradation is something that aggregators will need to worry about incorporating into their aggregation cost models when trying to maximize EV owner participation, making it of low concern for utilities. Rebound peaks, however, would prove costly for utilities and greatly affect the cost of electricity, meaning they are of high concern and would be a potential deterrent for utilities to partner with EV aggregators. Lastly, though no reported cyberattacks have occurred against US grids via EVSE, such an attack would have more dramatic impacts on grid-wide operations than at the EV owner level. Therefore, it is a moderate concern for utilities to implement the available cybersecurity measures in order to prevent attacks from interfering with grid operations.

\subsubsection{Aggregators}

As aggregators seek to maximize profit through trading energy, electricity price is of high concern with emissions ranging from low to moderate concern in order to account for potential monetary incentives for carbon-free electricity generation as discussed in \ref{subsubsec:Utilities}.

As discussed previously in Section 
 \ref{sec:Vulnerabilities}, battery degradation for EV owners can be accelerated by excessive use of V2G or certain charging schedule alterations. Because it is in aggregators' best interest to have high participation levels from EV owners \cite{AggEVImportance}, it should be a noteworthy concern for aggregators to create appropriate cost models that take this into account. Aggregators should also remain highly interested in avoiding rebound peaks in order to not deter utilities from cooperating with them. They should also be mindful of the potentially drastic consequences that could occur at the grid level from a scaled cyber attack. However, such an attack as noted previously would be unlikely so there will be only moderate concern with maintaining robust security measures. 

\subsubsection{Policy Makers}

Emissions should be of high priority for policymakers if they wish to influence the power sector to align more with government climate initiatives \cite{UNPolicyMakers}, which if left to its own devices, will trend towards the cheapest options regardless of environmental externalities \cite{EIADispatchOrder}. Electricity price is still listed as a moderate priority, as effective climate policy should attempt to not drastically increase electricity prices in order to be well received and sustainable \cite{EmissionsCostTrafeoff}.

Battery degradation is not easily or realistically controlled by policy, so it is of low concern for policymakers. Rebound peak is also difficult to control directly by policy, however excessive peaking events would require increased spending to maintain power systems infrastructure which may require public funding, so rebound peak is categorized as being of low to moderate concern for policymakers \cite{PublicAndFederalElectricity}. Lastly, though cyberattacks are currently scarce in the EV space, cyberattacks as a whole are becoming more common globally \cite{CyberAttackIncreases}. In light of this, the US is becoming increasingly concerned with cybersecurity policy \cite{USCyberPolicy}. Thus, it is categorized as moderate to high due to its increasing importance to policymakers.

\subsection{Conclusions}

EV aggregation has been gaining substantial attention and praise for its proven grid benefits regarding peak shaving, emission reductions, and electricity costs. Despite the benefits of effective EV aggregation, it's critical for many stakeholders to maintain awareness of the externalities that are inherent to the coordination of such a highly variable and widely distributed energy resource. Aggregation vulnerabilities, such as battery degradation, the creation of unintentional rebound peaks, and cybersecurity issues have the potential to offset some of the positive aspects of large-scale coordination of EV fleets. 

EV owners, utilities, aggregators, and policymakers are all uniquely impacted by such vulnerabilities, and understanding how different stakeholders will be concerned with different externalities will be crucial to their cooperation. These issues are also very likely to evolve going forward as new battery technologies emerge, electricity mixes change, and EV penetration rises, meaning that it will be paramount to maintain awareness of EV aggregation's ever-changing vulnerabilities in order to maximize the benefits of such a large, flexible, and mobile ancillary resource.

\IEEEoverridecommandlockouts
\maketitle

\bibliographystyle{IEEEtran}
\bibliography{ref.bib}

\end{document}